\documentclass[preprint,aps,amsmath,amsfonts,amssymb,superscriptaddress,showpacs]{revtex4}
\usepackage{epsfig}

\usepackage{graphics}
\usepackage{color} 
\usepackage{tikz} 

\usepackage{fancybox}
\usepackage{framed}

\begin{document}

\title{Record statistics for biased random walks, with an application to financial data}

\author{Gregor Wergen}
\affiliation{Institut f\"ur Theoretische Physik, Universit\"at zu K\"oln, 50937 K\"oln, Germany}
\author{Miro Bogner}
\affiliation{Institut f\"ur Theoretische Physik, Universit\"at zu K\"oln, 50937 K\"oln, Germany}
\author{Joachim Krug}
\affiliation{Institut f\"ur Theoretische Physik, Universit\"at zu K\"oln, 50937 K\"oln, Germany}



\begin{abstract} We consider the occurrence of record-breaking events in random walks with asymmetric jump distributions. The statistics of records in symmetric random walks was previously analyzed by Majumdar and Ziff
\cite{Majumdar} and is well understood. Unlike the case of symmetric jump distributions, in the asymmetric case the statistics of records depends on the choice of the jump distribution. We compute the record rate $P_n\left(c\right)$, 
defined as the probability for the $n$th value to be larger than all previous values, for a Gaussian jump distribution with standard deviation $\sigma$ that is shifted by a constant drift $c$. 
For small drift, in the sense of  $c/\sigma \ll n^{-1/2}$, the correction to $P_n\left(c\right)$ grows proportional to arctan$\left(\sqrt{n}\right)$ and saturates at the value $\frac{c}{\sqrt{2} \sigma}$. 
For large $n$ the record rate approaches a constant, which is approximately given by $1-\left(\sigma/\sqrt{2\pi}c\right)\textrm{exp}\left(-c^2/2\sigma^2\right)$ for $c/\sigma \gg 1$.  
These asymptotic results carry over to other continuous jump distributions with finite variance. As an application, we compare our analytical results to the record statistics of 366 daily stock prices from the Standard \& Poors 500  index. The biased random walk accounts quantitatively for the increase in the number of upper records due to the overall trend in the stock prices, and after detrending the number of upper records 
is in good agreement with the symmetric random walk. However the number of lower records in the detrended data is significantly reduced by a mechanism that remains to be identified.
\end{abstract}
\date{\today}
\maketitle

\section{Introduction}

The random walk is a paradigmatic model of statistical physics, which combines utmost conceptual simplicity with a surprising richness of emergent behaviors \cite{KRB2010,Weiss1994}. Among the many interesting features of random walks, recent research has focused in particular on its extremal properties, exploring quantities such as the height and position of the globally maximal excursion of a one-dimensional walk of a given number of steps, and the statistics of \textit{records} of this process \cite{Majumdar2}. Here a record is defined as an entry in a discrete, real valued time series that is larger (upper record) or smaller (lower record) than all previous entries. While the mathematical theory of records is well developed for time series of independent, identically distributed random variables \cite{Glick1,Arnold,Nevzorov}, little has been known about the record statistics of correlated processes. It is therefore remarkable that records of a large class of one-dimensional random walks can be characterized in considerable detail, as was shown in recent work by Majumdar and Ziff (MZ) \cite{Majumdar}. Specifically, they considered the random process defined by 
\begin{equation}\label{RW}
X_n = X_{n-1} + \xi_n,
\end{equation}
where $X_0 = 0$ (say) and the step sizes $\xi_n$ are independent, identically distributed random variables drawn from a probability density $\phi(\xi)$ that is required to be continuous and symmetric, but is otherwise arbitrary. We say that an upper record occurs at time $n$ if $X_n = \max\{X_0,...,X_n\}$. Based on the Sparre Andersen theorem for the survival probability of the random walk \cite{SparreAndersen,Feller,RednerBook,Godreche}, MZ show that the probability $\Pi\left(m,n\right)$ for the $n$th event to be the $m$th record is given by
\begin{eqnarray}\label{Pmnsymm}
 \Pi\left(m,n\right) = \left(2n-m+1\atop m\right)2^{-2n+m-1}
\end{eqnarray}
for $m \leq n+1$. The first moment of this distribution with respect to $m$ yields the mean number of records after $n$ steps, which equals  $m_n \approx \frac{2}{\sqrt{\pi}}\sqrt{n}$ for large $n$, and the probability $P_n$ for the $n$th event to be a record (henceforth referred to as the \textit{record rate}) decays like $P_n \approx \frac{1}{\sqrt{\pi n}}$. In the present paper we aim to generalize these results to random walks with asymmetric jump distributions. In the first part of the paper (Sections \ref{Section_survival} and \ref{Section_drift}) we study records generated by random walks with a symmetric jump distribution that have an additional constant \textit{drift} $c$, such that (\ref{RW}) generalizes to
\begin{equation}
\label{drift}
X_n = X_{n-1} + \xi_n + c
\end{equation}
with a symmetric jump distribution $\phi(\xi)$. For the special case of a Cauchy distribution this problem was considered previously in \cite{LeDoussal}. Here, we derive approximate results for the case of a Gaussian jump distribution that apply also more generally to distributions with a finite variance. 

Similar to our earlier work \cite{Franke2010,Wergen2010} on the related problem of records from independent random variables with drift \cite{LeDoussal,Ballerini}, our strategy will be to analyze the limiting cases of small and large drift, respectively, as quantified by the ratio $c/\sigma$ of the drift speed to the standard deviation $\sigma$ of the jump distribution $\phi(\xi)$.  For the Gaussian random walk we find that in the limit of $\frac{c}{\sigma} \ll \frac{1}{\sqrt{n}}$ the mean number of records and the record rate are given by 
\begin{eqnarray}
\label{Mnc} m_n\left(c\right) & \approx & \frac{2\sqrt{n}}{\sqrt{\pi}} + \frac{c}{\sigma}\frac{\sqrt{2}}{\pi}\left(n\; \textrm{arctan}\left(\sqrt{n}\right) - \sqrt{n}\right), \\
\label{Pnc} P_n\left(c\right) & \approx & \frac{1}{\sqrt{\pi n}} + \frac{c}{\sigma}\frac{\sqrt{2}}{\pi}\textrm{arctan}\left(\sqrt{n}\right).
\end{eqnarray}
In the limit of $\frac{c}{\sigma} \gg \frac{1}{\sqrt{n}}$ the record rate $P_n\left(c\right)$ approaches a constant value. If in addition $\frac{c}{\sigma}\gg1$, this constant is given approximately by
\begin{equation}
\label{limit}
\lim_{n \to \infty} P_n \approx
1-\frac{c}{\sqrt{2\pi}\sigma}e^{-\frac{c^2}{2\sigma^2}}.
\end{equation}

In Section \ref{Section_SP500} we apply our results to fluctuations in stock prices, arguably one of the most important  (and ancient) applications of random walk theory \cite{Bachelier,Mantegna2000,Voit2001}. The basic model of a stock price is the geometric random walk $S_n = e^{X_n}$ with an upward bias reflecting long-term economic growth. Our analysis of record events in the Standard \& Poors 500 index shows a corresponding surplus of upper record events, which is consistent with the theoretical expectation. However, an asymmetry between upper and lower records remains even when the bias has been (approximately) removed \cite{Bogner2009}, a feature that may be related to the gain-loss asymmetry reported in previous analyses of stock market fluctuations \cite{Jensen2003,Johansen2006,Simonsen2007,Karpio2007}. We conclude with a summary and a discussion of some open problems. 

\section{\label{Section_survival}Survival probabilities and First Passage Times}

The record statistics of a random walk can be analyzed by considering the generating functions of the survival and first passage probabilities of the process \cite{Majumdar,Majumdar2,LeDoussal}. In \cite{Majumdar} it was shown that the generating function of $\Pi\left(m,n\right)$ is of the form
\begin{eqnarray}\label{pmz}
 \sum_{n=m-1}^{\infty} \Pi\left(m,n\right) z^n = \tilde{f}_-^{m-1}\left(z\right)\tilde{q}_-\left(z\right),
\end{eqnarray}
where $\tilde{q}_\pm\left(z\right)$ is the generating function of the positive (negative) survival probability $q_\pm\left(n\right)$ of the random walk. The latter is defined as the probability that the process stays above (below) the origin up to the $n$th step. Similarly $\tilde{f}_\pm\left(z\right)$ is the generating function of the positive (negative) first-passage probability $f_\pm\left(n\right)$ of the random walk, with $f_\pm\left(n\right)=q_\pm\left(n-1\right)-q_\pm\left(n\right)$. In the case of the symmetric random walk considered in \cite{Majumdar} we have $q_-\left(n\right)=q_+\left(n\right)=q\left(n\right)$ and $f_-\left(n\right)=f_+\left(n\right)=f\left(n\right)$ and both $q\left(n\right)$ and $f\left(n\right)$ are completely universal for all continuous jump distributions.

Since we want to study asymmetric random walks, we need distinguish between positive and negative survival probabilities and first passage times, and consider the functions $q_\pm\left(n\right)$ and $f_\pm\left(n\right)$. As in \cite{Majumdar} a theorem by Sparre Andersen will play a key role in our considerations. In \cite{SparreAndersen, Godreche} it was shown that
\begin{eqnarray}\label{SparreAndersen}
 \tilde{q}_\pm\left(z\right) = \sum_{n=0}^\infty q_\pm\left(n\right)z^n = \textrm{exp}\left(\sum_{n=1}^\infty \frac{p_\pm\left(n\right)}{n}z^n\right),
\end{eqnarray}
where $p_\pm\left(n\right)$ is the probability for the walker to be above or below the origin at the $n$th step. This quantity can be easily computed from $p_\pm = \int_0^{\infty} G\left(\pm x,n\right)\mathrm{d}x$, where $G\left(x,n\right)$ is the positional probability density of a random walk of $n$ steps that started at the origin. Details on the computation of $G\left(\pm x,n\right)$ and $p_\pm\left(n\right)$ can be found in \cite{Majumdar2} and \cite{RednerBook}.  In the case of a symmetric random walk we simply have $p_\pm\left(n\right) = \frac{1}{2}$ independent of $n$ and we find that in this case $\tilde{q}_\pm\left(z\right) = \left(1-z\right)^{-\frac{1}{2}}$ and $q_\pm\left(n\right) = \left(2n\atop n\right)2^{-2n}$ \cite{Majumdar}. These results eventually lead to Eq. (\ref{Pmnsymm}) \cite{Majumdar}.

\begin{figure}
\centerline{\includegraphics[width=12.cm]{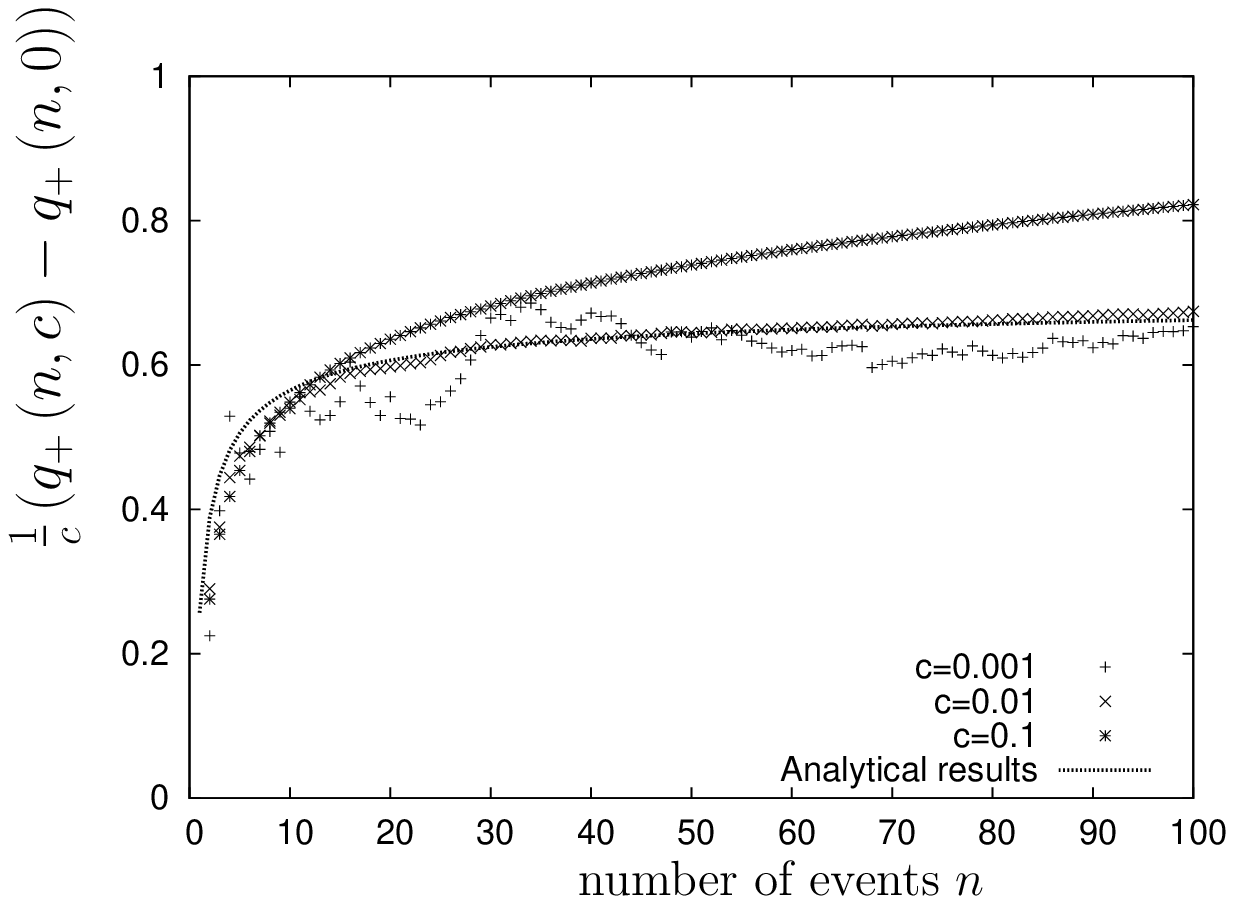}}
\caption{Relative effect of the drift on the positive survival probability $q_+\left(n,c\right)$ of a Gaussian random walk with $\sigma=1$. The effect of the drift is represented by $\frac{1}{c}\left(q_+\left(n,c\right)-q_+\left(n,0\right)\right)$ for different drift speeds $c$. We simulated $10^7$ realization of a random walk with $n=100$ steps for each drift speed. The dotted line represents the analytical results obtained in Eq.~(\ref{qn}). For small drift $c=0.001$ and $c=0.01$ we find good agreement with this approximation. }
\label{Fig:Qnc}
\end{figure}

In the case of an asymmetric random walk the situation gets a bit more complicated. We compute $q_\pm\left(n\right)$ and its generating function for a Gaussian random walk with drift $c$. Here, the jump distribution 
of the symmetric random variable $\xi$ in (\ref{drift}) is of the form
$ \phi\left(\xi\right) = \frac{1}{\sqrt{2\pi\sigma}}e^{-\frac{\xi^2}{2\sigma^2}}$
with standard deviation $\sigma$. It is easy to show that the probability density $G\left(x,n\right)$ of the random walk after $n$ steps is given by $G\left(x,n\right) = \frac{1}{\sqrt{2\pi n}} \textrm{exp}\left(-\frac{\left(x-nc\right)^2}{2\sigma^2 n}\right)$ and $p_\pm\left(n\right) = \frac{1}{2}\left(1\pm\textrm{erf}\left(\sqrt{\frac{n}{2}}\frac{c}{\sigma}\right)\right)$. We start with the case of a small linear drift with $c\ll\frac{\sigma}{\sqrt{n}}$ such that $p_\pm\left(n\right)\approx \frac{1}{2} \pm \sqrt{\frac{n}{2\pi}}\frac{c}{\sigma}$. Now we can employ Eq. (\ref{SparreAndersen}). Expanding up to first order in $c$ we find
\begin{eqnarray}\label{qz}
\tilde{q}_\pm\left(z\right) \approx \frac{1}{\sqrt{1-z}}\left(1 \pm \frac{c}{\sqrt{2\pi}\sigma} \sum_{n=1}^{\infty} \frac{z^n}{\sqrt{n}}\right).
\end{eqnarray}
With $\sqrt{1-z}^{-1} = \sum_{n=0}^{\infty} \left(2n\atop n\right)2^{-2n}z^n$ and making use of the Cauchy formula for products of infinite sums we obtain the following expression for $\tilde{q}_\pm\left(z\right)$:
\begin{eqnarray}
 \tilde{q}_\pm\left(z\right) \approx \frac{1}{\sqrt{1-z}} \pm \frac{c}{\sqrt{2\pi}\sigma}\sum_{n=0}^\infty \sum_{k=0}^n \left(2k\atop k\right) \frac{2^{-2k} z^{n+1}}{\sqrt{n-k+1}}.
\end{eqnarray}
The binomial coefficient can be approximated by $\left(2k\atop k\right) \approx 4^k / \sqrt{\pi k}$ and with this we can approximate the sum over $k$ by an integral,
\begin{eqnarray} 
 \sum_{k=0}^n \frac{1}{\sqrt{\pi k}} \frac{z^{n+1}}{\sqrt{n-k+1}} \approx \frac{1}{\sqrt{\pi}} \int_0^n \frac{\mathrm{d}k z^{n+1}}{\sqrt{k} \sqrt{n-k+1}} \approx \sqrt{\pi} z^{n+1}.
\end{eqnarray}
We thus obtain a simple result for the generating function of the survival probability,
\begin{eqnarray}
 \tilde{q}_\pm\left(z\right) \approx \frac{1}{\sqrt{1-z}} \pm \frac{c}{\sqrt{2}\sigma} \sum_{n=1}^\infty z^n,
\end{eqnarray}
and finally the following expression for the survival probability $q_\pm \left(n\right)$ under a small linear drift:
\begin{eqnarray}\label{qn}
q_\pm\left(n\right) \approx \frac{1}{\sqrt{\pi n}} \pm \frac{c}{\sqrt{2}\sigma}.
\end{eqnarray}
The first term on the right hand side is the result for the symmetric random walk discussed in \cite{Majumdar}, which is now supplemented by a correction linear in $\frac{c}{\sigma}$. Although this particular result for $q_\pm\left(n\right)$ will not be needed in our derivation of the record statistics, we found it useful to test it against numerical simulations. The results are shown in Fig.~\ref{Fig:Qnc}. For small $c$, Eq.(\ref{qn}) is in good agreement with the simulations. 

For the sake of completeness we also provide the small $c$ expansion of $\tilde{f}_\pm\left(z\right)$ that will become important later. With $\tilde{f}_\pm\left(z\right) = 1-\left(1-z\right)q_\pm\left(z\right)$ we find 
\begin{eqnarray}\label{fz}
\tilde{f}_\pm\left(z\right) \approx 1-\sqrt{1-z}\left(1 \pm \frac{c}{\sqrt{2\pi}\sigma}\sum_{n=1}^{\infty}\frac{z^n}{\sqrt{n}}\right).
\end{eqnarray}
From this we obtain, by methods very similar to those used above to derive $q_\pm\left(n\right)$, the result
\begin{eqnarray}
f_\pm\left(n\right) \approx \frac{1}{2\sqrt{\pi}} n^{-\frac{3}{2}} \pm \frac{c}{\sqrt{2}\pi\sigma} n^{-\frac{1}{2}}.
\end{eqnarray}

Next we consider the case of large drift, $\frac{c}{\sigma} \gg 1$. Here we will only discuss $q_-\left(n\right)$, as this is the quantity needed for the computation of the record rate; $q_+\left(n\right)$ has a different behavior in this regime. In the limit of $\frac{c}{\sigma} \gg 1$ we find that $p_-\left(n\right) \approx \sigma\left(2\pi n c^2\right)^{-1/2}e^{-\frac{c^2 n}{2\sigma^2}}$. Using this we find that for large $n$, $\tilde{q}_-\left(z\right)$ and $q_-\left(n\right)$ are of the form
\begin{eqnarray}\label{qzlargec}
\tilde{q}_-\left(z\right) & \approx & 1 + \sum_{n=1}^\infty \frac{\sigma}{c \sqrt{2\pi n^3}}e^{-\frac{c^2 n}{2\sigma^2}} z^n,\\ 
q_-\left(n\right) & \approx & \frac{\sigma}{c\sqrt{2\pi n^3}} e^{-\frac{c^2 n}{2\sigma^2}}.
\end{eqnarray}
These particular results were already reported in \cite{LeDoussal}. At this point it is important to notice that all results concerning the first-passage and survival probabilities in the large $n$ limit are easily transferable to other jump distributions as long as these have a finite variance. Because of the central limit theorem, $G\left(\pm x,n\right)$ and therefore $p_\pm\left(n\right)$ will approach the same expressions for large $n$ as were derived here for the Gaussian jump distribution.

\section{\label{Section_drift}Gaussian random walks with drift}

\subsection{Record rate for small $c/\sigma$ and $n \ll (\sigma/c)^2$}

With the small $c$ expansions in Eqs.~(\ref{qz}) and (\ref{fz}) we have all ingredients needed to derive the record statistics for a Gaussian random walk with a small linear drift. We start by computing the mean number of records $m_n$ expected up to the $n$th step. For the generating function $\tilde m(z) = \sum_{n=0}^\infty m_n z^n$ of this quantity it was found in \cite{LeDoussal} that $\tilde{m} \left(z\right) = 1/\left(\left(1-z\right)^2 \tilde{q}_-\left(z\right)\right)$, a result that can be obtained by computing the first moment of Eq.~(\ref{pmz}). We can now evaluate this expression making use of the generating function for $q_-\left(n\right)$ given in Eq.(\ref{qz}). In the limit of small $\frac{c}{\sigma}$ this yields
\begin{eqnarray}
\tilde{m}\left(z\right) \approx \frac{1}{\sqrt{1-z}^3}\left(1+\frac{c}{\sqrt{2\pi}\sigma}\sum_{n=1}^{\infty} \frac{z^n}{\sqrt{n}}\right).
\end{eqnarray}
Using the series expansion of $\sqrt{1-z}^{-3}$ and employing once again the Cauchy formula for infinite sums and the Stirling approximation, we find
\begin{eqnarray}
\tilde{m}\left(z\right) \approx \frac{1}{\sqrt{1-z}^3} + \frac{\sqrt{2}c}{\pi\sigma} \sum_{n=1}^{\infty} z^n \sum_{k=0}^{n-1} \frac{\sqrt{k}}{\sqrt{n-k+1}}.
\end{eqnarray}
If $n$ is not too small, the sum over $k$ can be replaced by an integral and we finally obtain an approximate expression for the generating function of $m_n$,
\begin{eqnarray}
\tilde{m}\left(z\right) \approx \frac{1}{\sqrt{1-z}^3} + \frac{\sqrt{2}c}{\pi\sigma} \sum_{n=1}^{\infty} z^n \left(n\; \textrm{arctan}\left(\sqrt{n}\right)-\sqrt{n}\right).
\end{eqnarray}
\begin{figure}
\centerline{\includegraphics[width=12.cm]{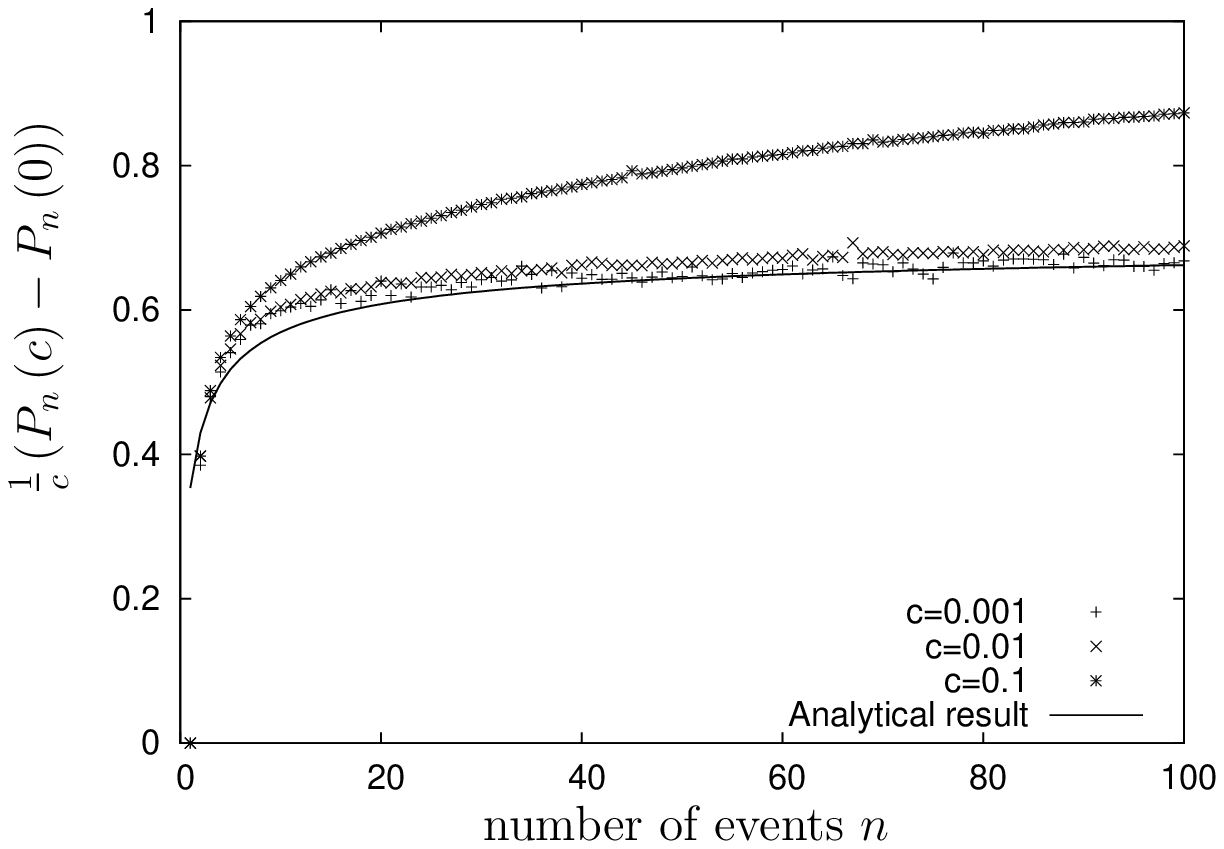}}
\caption{Relative effect of the drift on the record rate $P_n\left(c\right)$ of a random walk with a Gaussian jump distribution ($\sigma=1$). The effect is represented by $\frac{1}{c}\left(P_n\left(c\right)-P_n\left(0\right)\right)$ for different drift speeds $c$. Again, we simulated $10^7$ realizations of a random walk with $n=100$ steps for each drift speed. The line represents the analytical results obtained in Eq.~(\ref{Pnc}). For small drift speeds $c=0.001$ and $c=0.01$ we find good agreement with the approximation, but for $c=0.1$ the approximation is no longer accurate. }
\label{Fig:Pnc}
\end{figure}
\begin{figure}
\centerline{\includegraphics[width=12.cm]{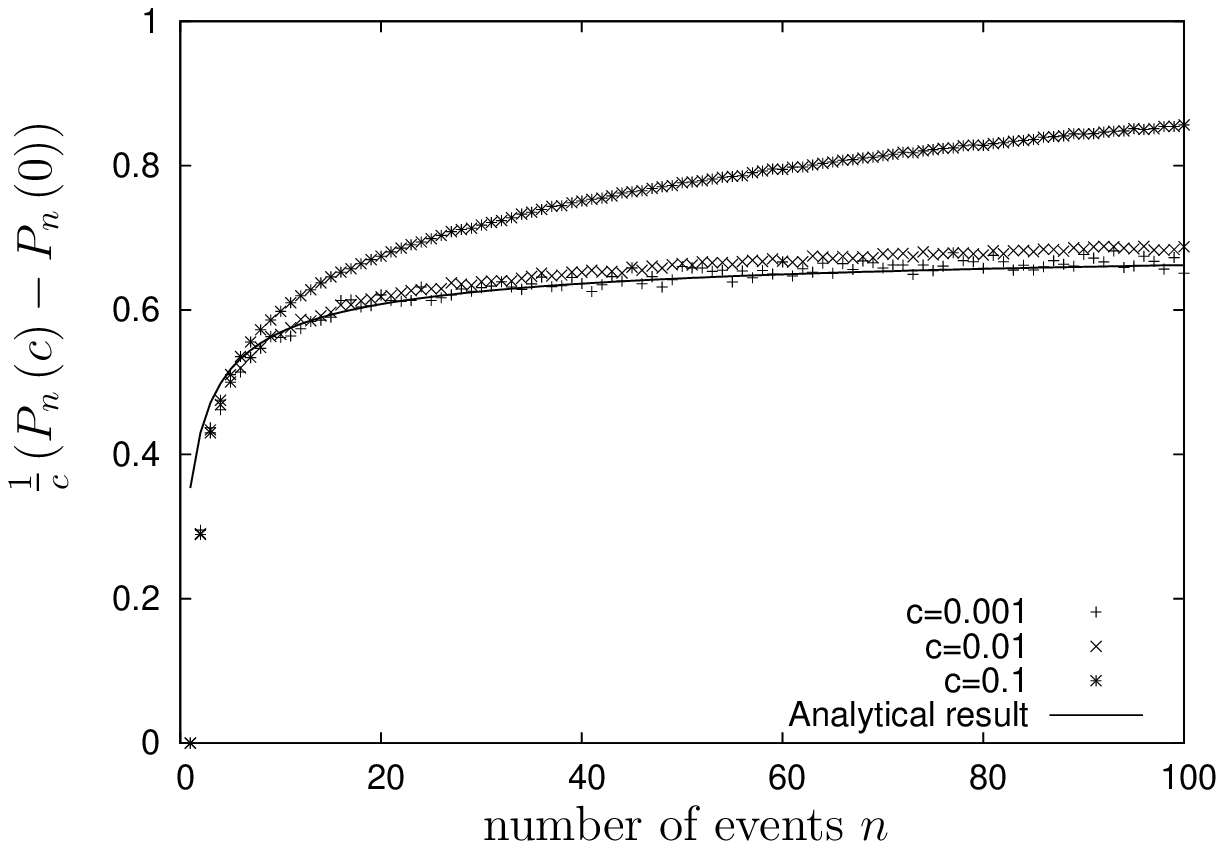}}
\caption{Relative effect of the drift on the record rate $P_n\left(c\right)$ for a random walk with a uniform jump distribution with standard deviation $\sigma=1$. The parameters of the simulation are the same as in Fig. \ref{Fig:Pnc}. Even though the expression (\ref{Pnc}) was derived for a Gaussian jump distribution, it is in a good agreement with the numerical results for small $c$.}
\label{Fig:Pncu}
\end{figure}
The mean number of records of the random walk with a small linear drift $c$ is therefore approximately given by
\begin{eqnarray}
 m_n \approx \left(2n \atop n\right) \frac{2n+1}{2^{2n}} + \frac{\sqrt{2}c}{\pi\sigma}\left(n\; \textrm{arctan}\left(\sqrt{n}\right) - \sqrt{n}\right).
\end{eqnarray}
Making use of the Stirling approximation this yields the previously announced expression (\ref{Mnc}) for $m_n(c)$ and, by taking a derivative with respect to $n$, the record rate $P_n\left(c\right)$ in the large $n$ limit 
as given in Eq.(\ref{Pnc}). The leading order correction  of the record rate due to the drift is seen to increase with $\textrm{arctan}\left(\sqrt{n}\right)$ and for larger $n$ it approaches a constant value. 
For large $n$ (but still in the regime $\frac{c}{\sigma}\ll\frac{1}{\sqrt{n}}$) we find the simple result 
\begin{equation}
\label{Pnsimple}
P_n\left(c\right) \approx \frac{1}{\sqrt{\pi n}} + \frac{c}{\sqrt{2}\sigma}. 
\end{equation}
We compared Eq.(\ref{Pnc}) to simulations and found good agreement in the regime $\frac{c}{\sigma}\ll\frac{1}{\sqrt{n}}$ (Fig. \ref{Fig:Pnc}). We also compared this result with numerical simulations of the record rate for random
walks with step sizes drawn from a uniform distribution (Fig. \ref{Fig:Pncu}). The results for the Gaussian and the uniform distribution are very similar to each other already for small $n$, reflecting the convergence
expected from the central limit theorem. 

\begin{figure}
\includegraphics[width=12.cm]{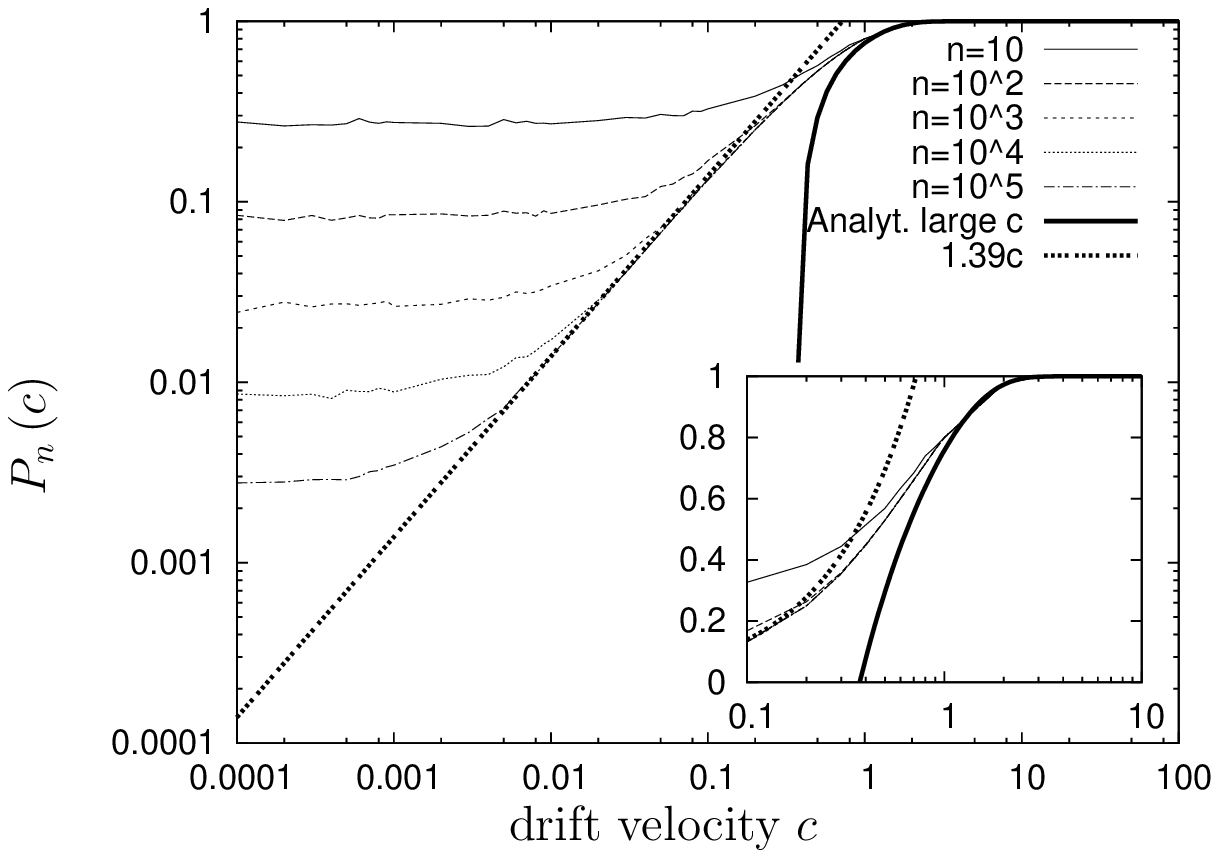}
\caption{Record rate for a biased Gaussian random walk with standard deviation $\sigma = 1$. The figure illustrates the convergence of $P_n(c)$ to the asymptotically constant record rate $P(c)$ for $n \to \infty$.
The inset shows that the large drift result (\ref{Pnlargec}) becomes accurate for $c/\sigma > 1$, and the bold dotted line shows that $P(c) \approx 1.39 \frac{c}{\sigma}$ for $c \to 0$.}
\label{Fig:Pnclargen}
\end{figure}

\begin{figure}
\includegraphics[width=12.cm]{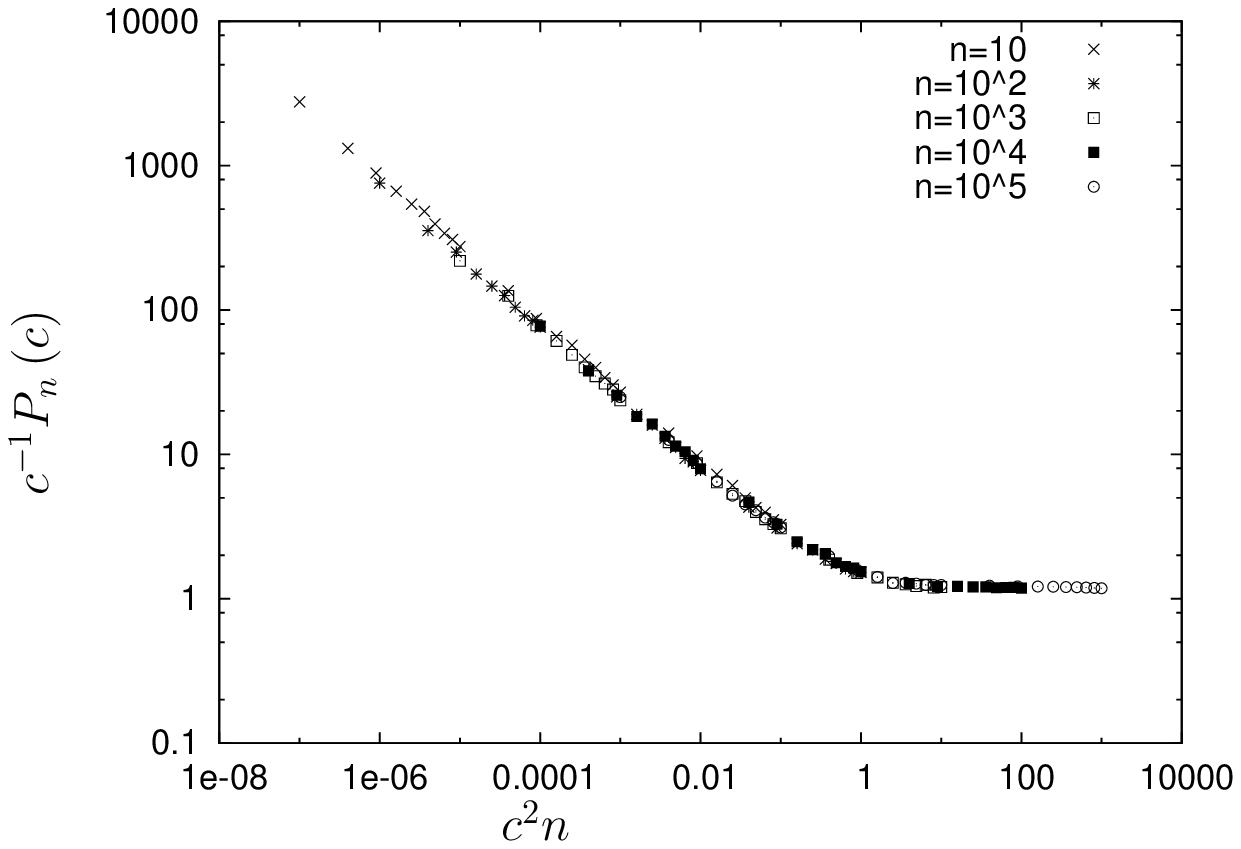}
\caption{Illustration of the conjectured scaling collapse (\ref{Pnscal}) of the record rate $P_n(c)$ for Gaussian random walks with $\sigma = 1$ and various drift speeds 
$c \leq 0.1$.}
\label{Fig:Pncscal}
\end{figure}

\subsection{Asymptotic record rate for large $n$}

Next we consider the limit of strong drift, $\frac{c}{\sigma} \gg 1$. Applying the same method as above and making use of our result (\ref{qzlargec}) for $\tilde{q}_-\left(z\right)$ in the regime
of large $c/\sigma$, we find that
the number of records increases linearly with time according to 
\begin{eqnarray}
\label{mnclarge}
 m_n(c) \approx n\left(1-\frac{\sigma}{\sqrt{2\pi}c} e^{-\frac{c^2}{2\sigma^2}}\right).
\end{eqnarray}
Correspondingly the record rate $P_n$ is independent of $n$ in this case. In fact, simulations show that the record rate approaches a finite, nonzero limit $P(c) \equiv \lim_{n \to \infty} P_n(c)$ for 
$n \to \infty$ for any positive value of the drift (Fig. \ref{Fig:Pnclargen}). This can be understood, on the basis of the general relation (\ref{pmz}) between the distribution of record events
and the negative first passage probability, to be a consequence of the fact that the negative mean first passage time of a random walk with positive drift is finite \cite{Feller,RednerBook}; roughly speaking,
one expects that the asymptotic record rate $P(c)$ is proportional to the inverse of the negative mean first passage time. The result (\ref{mnclarge}) implies that the asymptotic record rate behaves as
\begin{eqnarray}\label{Pnlargec}
 P \left(c\right) \approx 1-\frac{\sigma}{\sqrt{2\pi}c} e^{-\frac{c^2}{2\sigma^2}}
\end{eqnarray}
for large $c/\sigma$ (see inset of Fig.\ref{Fig:Pnclargen}). Furthermore, since the negative mean first passage time diverges as $c^{-1}$ for $c \to 0$ \cite{Feller,RednerBook}, 
the asymptotic record rate should behave as $P(c) \sim c$ for small $c$.
This is confirmed by the simulations, which indicate that $P(c) \approx 1.39 \; (c/\sigma)$ for $c/\sigma \ll 1$.

The time scale $n^\ast(c)$ at which the saturation of the record rate occurs can be estimated by comparing the two terms in Eq.(\ref{Pnsimple}), which shows that 
\begin{equation}
\label{nast}
n^\ast \sim \left(\frac{\sigma}{c}\right)^2 
\end{equation}
for small $c$. Not surprisingly, this is also the time scale at which the drift begins to dominate the mean square displacement of the random walk. 
Together with the linear behavior of the asymptotic record rate, this suggests the scaling form
\begin{equation}
\label{Pnscal}
P_n(c)  = \frac{c}{\sigma} g((c/\sigma)^2 n)
\end{equation}
for small $c/\sigma$ and arbitrary $n$, where the limiting behaviors of the scaling function are $g(x \to 0) \approx \frac{1}{\sqrt{\pi x}}$ and 
$g(x \to \infty) \approx 1.39$. This relation is well fulfilled by the numerical data shown in Fig.\ref{Fig:Pncscal}.

\section{\label{Section_SP500}Record statistics of stock prices in the S\&P 500}

A prominent application of the random walk process can be found in the financial sciences. Originally introduced by Bachelier in 1900 \cite{Bachelier}, the geometric random walk is the standard model used to 
describe the evolution of stock prices. In the application of this model to actual data, 
trends are always an issue, which in the simplest case are described by a linear drift in the logarithm of the stock price. 
In this section we present an empirical analysis of record events 
in historical stock prices taken from the Standard \& Poors 500 index, and compare the results to the theoretical predictions derived above. 
The observational data we used consist of daily recordings of 366 stocks that were contained in the index from January 1990 to March 2009, resulting in 366 time series of length $n=5000$ \cite{SPdata}. 
We first analyzed the recordings without any detrending and then considered detrended data in which a fitted linear trend was subtracted from the logarithms of the stock prices. 

\begin{figure}
\includegraphics[width=12.cm]{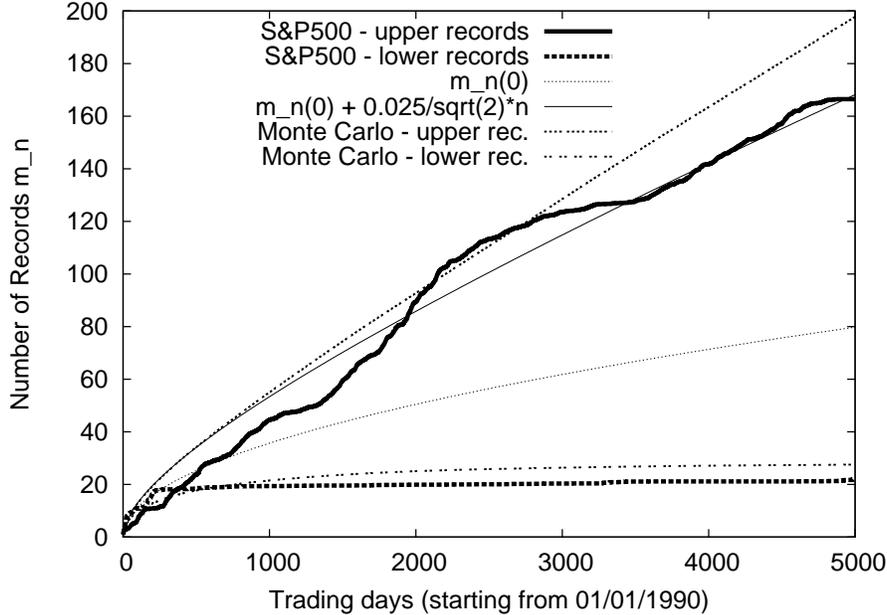}
\caption{Mean number of records averaged over  $366$ stocks from the S\&P 500 index, computed from daily va1ues for the time period 1.1.1990 - 31.3.2009. Full thick line shows the number of upper records, dotted thick line shows the number of lower records in the data set. The expected number of records $m_n\left(0\right) = \frac{2}{\sqrt{\pi}} \sqrt{n}$ for a symmetric random walk is shown by the thin dotted line.  
Also shown are the predictions of the biased random walk model with effective normalized drift 
$c/\sigma = 0.025$ obtained from Monte Carlo simulations as well as from the approximate expression $m_n(c) = m_n(0) + \frac{c}{\sqrt{2}\sigma} n$ (thin full line).}
\label{Fig:sp1}
\end{figure}

In the raw stock data the number of upper records after $n=5000$ trading days is considerably larger than the expected number of $2\sqrt{5000/\pi} \approx 79.79$ for a symmetric random walk. At the end of the observation
period, we found an average number of $166.56$ upper records in the stocks, but only $22.33$ lower records. The rate of upper records was roughly constant over the entire period, whereas the rate of lower records was almost zero already after 300 days. Apparently a positive trend had a very strong effect on the record statistics of the analyzed stocks. 
To quantify the trend, we performed a linear regression analysis on the logarithms of the individual stock prices, determining the drift $c_i$ and the standard deviation of increments $\sigma_i$ for each stock
$i=1,...,366$. The normalized drift $c_i/\sigma_i$ was then averaged over all stocks, yielding the estimate  
$\langle c_i/\sigma_i \rangle \approx 0.025$. At $n=5000$ we are thus well outside the regime in which the pertubative result (\ref{Mnc}) should be valid. Still, inserting the estimated
normalized drift $c/\sigma = 0.025$ into (\ref{Mnc}) we obtain a record number of 166.59, in very close agreement with the observed value. The comparison with 
Monte Carlo simulations of biased random walks with the same drift shows that this accuracy is actually fortuitous, but the description of the stock market data
by the biased random walk model is nevertheless quite reasonable (Fig. \ref{Fig:sp1}).

\begin{figure}
\includegraphics[width=12.cm]{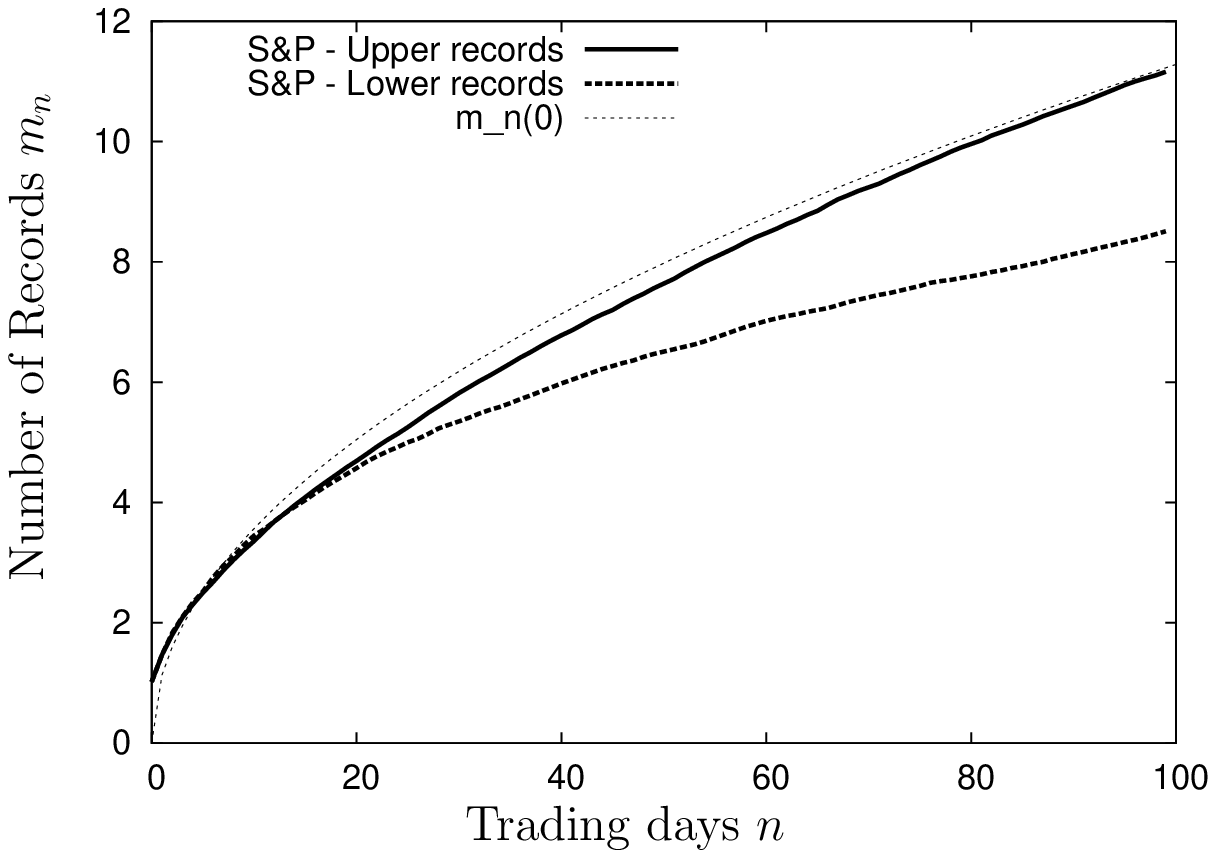}
\caption{Mean number of records in subsequences of the time series taken from the S\&P 500 index. The entire data set of 5000 consecutive daily values was split into 50 subsequences of length 100. For each of the subsequences a linear detrending of the logarithm of the daily values was performed and the upper and lower record numbers were determined from the detrended data. The results, averaged over all stocks and all subsequences,  are given by the thick black line (upper records) and the thick dashed line (lower records). The thin dashed line shows the analytical prediction for a symmetric random walk $m_n\left(0\right) = 2\sqrt{n/\pi}$. The number of upper records is in good agreement with $m_n\left(0\right)$, but the number of lower records is significantly reduced.}
\label{Fig:sp2}
\end{figure}

Next we detrended the data by subtracting the fitted linear trend from the logarithmic stock prices, and counted the number of records in the detrended time series. 
We found an average number of $75.79$ upper records after $5000$ steps, in close agreement with the result for a symmetric random walk. However, the number of lower records was only $53.65$, which is significantly smaller
than expected. This residual asymmetry between upper and lower records persists if, instead of subtracting an overall linear trend, the data are detrended by normalizing each stock by the index \cite{Bogner2009}. 
To further explore this phenomenon we split the time series into $50$ shorter series each lasting $100$ trading days. We detrended each of the shorter time series individually by subtracting a linear trend, counted the number 
of upper and lower records, and then averaged the record numbers over the whole ensemble of $50 \times 366$ series of length 100. The results are shown in Fig. \ref{Fig:sp2}. 
It appears that while the number of upper records is in a very good agreement with the symmetric random walk model, the number of lower records is still suppressed. This effect was found for different choices 
of the lengths of the time series and appears to be independent of this choice. 

Qualitatively, a reduced number of lower records indicates that the positive first-passage times are increased compared to the 
corresponding negative first passage times. An asymmetry between first passage times to a prescribed (positive or negative) return level has in fact been observed 
in previous analyses of stock market data, and is known as the \textit{gain-loss asymmetry} \cite{Jensen2003,Johansen2006,Simonsen2007,Karpio2007}. 
However, this phenomenon differs in several important respects from the one reported here. First, in most (though not all \cite{Karpio2007}) cases the sign of the asymmetry is opposite to that
suggested by the asymmetry in the record statistics, in that first passage times for crossing a prescribed level from below are larger than for crossings from above \cite{Jensen2003,Simonsen2007}. Second,
the asymmetry vanishes when the prescribed return level tends to zero, which is the relevant limit for the analysis of records. Finally, in contrast to the asymmetry between upper
and lower records reported here, the gain-loss asymmetry is a property of entire stock indices which does not occur in individual stocks \cite{Johansen2006,Simonsen2007}.
Indeed, a preliminary analysis of first-passage times to the origin in the detrended S\&P 500 data shows an asymmetry between positive and negative excursions only when 
the starting point of the excursion is conditioned to be a record event \cite{Wergen}.  
An explanation of the observed residual asymmetry between upper and lower records must therefore
be left to future work.

\section{Summary} In conclusion, using the methods introduced in \cite{Majumdar} and a more general form of the Sparre Andersen Theorem \cite{SparreAndersen,Majumdar2}, we were able to describe the effect of a linear drift on the record statistics of a Gaussian random walk in two regimes. For short times $n \ll \left(\frac{\sigma}{c}\right)^2$ we find that the correction to the record rate $P_n\left(c\right)-P_n\left(0\right)$ increases proportional $\textrm{arctan}\left(n\right)$ and then saturates at a value of $\frac{c}{\sqrt{2}\sigma}$. On the other hand, for large $n$ the record rate saturates at a constant limiting value $P(c)$, which is linear in $c$ for 
$c/\sigma \ll 1$ and approaches unity for large $c/\sigma$ according to  Eq.(\ref{Pnlargec}). The transition between the two regimes is described by the scaling form (\ref{Pnscal}). 

We applied our results to the statistics of records in $366$ stocks contained in the S\&P 500 index from 1990 to 2009. We found that, after detrending, the number of upper records in the stocks is basically identical to that predicted for the symmetric random walk. The fact that the number of lower records appears to be systematically decreased is interesting and needs to be examined more thouroughly in the future.
On the theoretical side, a possible topic for 
future research is the record statistics of asymmetric random walks with a more complicated asymmetry than just a constant drift. 
The issue of asymmetric random walks with discrete jump distributions is also still open for further investigations.

\section*{Acknowledgements} We thank I. Szendro and J. Franke for helpful discussions and
support, and gratefully acknowledge the Department of Business Administration and
Finance at the University of Cologne for providing access to the stock data from the S\&P 500. Financial support was provided 
by DFG within the Bonn Cologne Graduate School of Physics and Astronomy.

\end{document}